\documentclass[reprint,twocolumn,preprintnumbers,pre,amsmath,superscriptaddress,amssymb]{revtex4}\usepackage{float} 
\usepackage{graphicx}
\usepackage{dcolumn}
\usepackage{bm}
\usepackage{xcolor}

\begin{document}

\title{Modal complexity as a metric for Anderson localization}
\author{Sandip Mondal}
\affiliation{Nano-optics and Mesoscopic Optics Laboratory, Tata Institute of Fundamental Research, 1, Homi Bhabha Road, Mumbai, 400005, India}
\author{Kedar Khare}
\affiliation{Optics and Photonics Centre, Indian Institute of Technology Delhi, New Delhi 110016, India }
\author{Sergey E. Skipetrov}
\affiliation{Univ. Grenoble Alpes, CNRS, LPMMC, 38000 Grenoble, France}
\author{Martin Kamp}
\affiliation{Julius-Maximilians-Universit\"{a}t Würzburg, Physikalisches Institut, Lehrstuhl f\"{u}r Technische Physik, Am Hubland, 97074 W\"{u}rzburg, Deutschland}

\author{Sushil Mujumdar}
\email[]{mujumdar@tifr.res.in}
\homepage[]{http://www.tifr.res.in/~mujumdar}
\affiliation{Nano-optics and Mesoscopic Optics Laboratory, Tata Institute of Fundamental Research, 1, Homi Bhabha Road, Mumbai, 400005, India}

\begin{abstract}
We present a thorough study of the complexity of optical localized modes in two-dimensional disordered photonic crystals. Direct experimental measurements of complexity were made using an interferometric setup that allowed for extraction of phases and, hence, complex-valued wavefunctions. The comparison of experimental and theoretical results allows us to propose a metric for Anderson localization based on the average value and statistical distribution of complexity. Being an alternative to other known criteria of localization, the proposed metric exploits the openness of the disordered medium and provides a quantitative characterization of the degree of localization allowing for determining the localization length. 

\end{abstract}

\pacs{Valid PACS appear here}
\maketitle

The mathematically rigorous definition of Anderson localization (AL) \cite{anderson1958, sheng1995, lee1985} requires exponential decay of the eigenstate intensity $|\psi(\mathbf{r})|^2$ in a disordered medium with the distance from its ``localization center'' $\mathbf{r}_c$: $|\psi(\mathbf{r})|^2 \leq a \exp(-b |\mathbf{r}-\mathbf{r}_c|)$ for sufficiently large $|\mathbf{r}-\mathbf{r}_c|$ and with positive $a$, $b$.
AL is expected to take place when the scattering mean free path $\ell$ becomes of order or shorter than the wavelength $\lambda$, which is known as the Ioffe-Regel criterion of localization \cite{ioffe1960, laurent2007,skipetrov2018}. Both the exponential decay of $|\psi(\mathbf{r})|^2$ and the condition $\ell \lesssim \lambda$ are difficult to use in experimental and numerical studies of AL because $\ell$ is not easy to determine and the investigated disordered samples are always of finite extent, so that, sufficiently large distances $|\mathbf{r}-\mathbf{r}_c|$ may not be accessible. To circumvent this difficulty, Edwards and Thouless proposed a localization criterion that \textit{exploits} the finite size and the openness of a sample: the dimensionless conductance of the sample $g$ should be small ($g < g_c \sim 1$) if the states $\psi(\mathbf{r})$ are spatially localized and large ($g > g_c$) otherwise \cite{edwards1972,thouless1977}. The scaling theory of localization relaxed the requirement of knowing $g_c$ by moving the focus on the evolution of $g$ with the sample size $L$ rather than on its precise value: the exponential decrease of $g$ with $L$ in the regime of AL is contrasted with its power-law growth for extended states \cite{abrahams1979}. Despite the scaling theory being a milestone in our understanding of AL and its major role in theoretical studies of AL \cite{evers2008}, it remains difficult to use this theory to interpret experiments because this would require fabrication of a series of disordered samples of different sizes but with exactly the same statistical properties of disorder. As a result, other less rigorous but easier-to-apply criteria of AL have been used in recent experiments and numerical simulations: enhanced fluctuations of the scattered intensity \cite{chabanov2000,hu2008,riboli2011,yamilov2023}, non-exponential decay of the time-of-flight distribution in transmission \cite{hu2008,storzer2006,yamilov2023}, and arrest of transverse expansion of a transmitted beam \cite{hu2008,sperling2013,yamilov2023}.


In this Letter, we propose and experimentally test a new metric of AL that exploits the finite size and the openness of the disordered sample in the spirit of Thouless criterion, and relies on the statistical properties of wave field fluctuations beyond intensity statistics. The criterion is based on the probability distribution $P(q^2)$ and the average value $\langle q^2 \rangle$ of the complexity $q^2 = \int \psi_i^2({\bf r}) d{\bf r}/\int \psi_r^2({\bf r}) d{\bf r}$ of the wave function $\psi(\mathbf{r}) = \psi_r(\mathbf{r}) + i\psi_i(\mathbf{r})$. In general, $q^2$ characterizes the transition from real/static wavefunctions $\psi({\bf r})$ to complex/propagating ones due to the the non-Hermiticity of the considered physical system \cite{langen1997, tzortzakakis2021, balasubrahmaniyam2020}. It was extensively studied in open chaotic billiards \cite{langen1997,brouwer2003,poli2009} but very few works exist for disordered media \cite{vanneste2009}.
With increasing disorder, Anderson localization effects come into play  and suppress the coupling of modes to the outside world, thereby inducing a certain degree of mode stationarity. This observation allows us to propose the complexity $q^2$ as a measure of degree of localization. As we demonstrate below, such an approach is a viable alternative to using Ioffe-Regel or Thouless critera of localization.
Its main advantage is that it requires neither measuring $k = 2\pi/\lambda$ and $\ell$ (which, in addition, may be loosely defined when Anderson localization sets in \cite{skipetrov2018}) nor relying on arbitrary critical values of $k\ell$ or $g$, both of order 1 but unknown otherwise. Compared to the direct application of scaling theory, using complexity eliminates the need for varying the sample size $L$ in order to determine the dependence $g(L)$, which is a great simplification for most of experiments. In addition, the proposed approach allows for quantitative characterization of localized states giving access to the localization length $\xi$.

To demonstrate the use of complexity to reveal and characterize Anderson localization, we perform optical experiments in two-dimensional (2D) disordered photonic crystals with open boundaries.
Previously, disordered photonic crystal samples were employed for studying various aspects of one-dimensional Anderson localization within PC waveguides.  \cite{sapienza2010, garcia2010, javadi2014, faggiani2016}. We address two-dimensional localization over deliberately disordered lattices, as described further. We produce our samples by lithographically writing air holes on thin GaAs membranes (thickness $340$~nm). A triangular lattice (lattice constant $a=630$~nm and radius of holes $r=139$~nm) is used as the underlying periodic structure. The disordered samples are made by deliberately displacing each hole from its original lattice site. See Supplementary Information for details.  
Structure factor calculations reveal no short range correlations in our disordered samples at any disorder \cite{mondal2019, kumar2020}. Two representative scanning electron microscope (SEM) images of crystalline and disordered samples are shown in Figs. 1(a) and (b).
The overall sample dimensions are $20\mu$m$\times20\mu$m and the experiment is performed in the near infrared regime ($\lambda\sim1550$ nm). Phase measurements are implemented through an interferometric setup, as illustrated in Fig~1(c). 
We use the interferometric phase measurement technique that is well-documented in the literature  \cite{daryanoosh2018, takeda2012, wolley2023}, and adopt the Fourier transform method (FTM).
FTM involves Fourier-space filtering of the recorded interference pattern in order to isolate the cross-term of interest as explained later. Further, we employ an efficient computational method which allows us to determine the carrier-fringe frequency to a sub-pixel accuracy, so that, the measured phase map is free of any undesirable tilt phase background \cite{singh2016, goyal2024}. This technique allows rapid high-fidelity phase retrieval and makes it possible to measure optical wavefunctions over several configurations and various degrees of disorder in a unified manner.

Light output from a tunable NIR diode laser is tightly focused onto the edge of the free-standing GaAs membrane. 
While the light is essentially transversely confined within the membrane, the inherent surface roughness allows a fraction of the light to scatter out of the plane. Owing to the miniscule roughness, this out-of-plane scatter does not alter the mode structure, and hence acts as a non-invasive probe of the modes. The scattered light is imaged onto an InGaAs CCD using a 100X objective, and is made to interfere with the reference beam, as shown. The mode structure, in absence of the reference beam, is hereafter referred to as the `object' beam. At every modal wavelength, three images corresponding to the object, reference and their interference are recorded and used to calculate the phase as discussed below.
\begin{figure}
	\includegraphics[width=8.5cm]{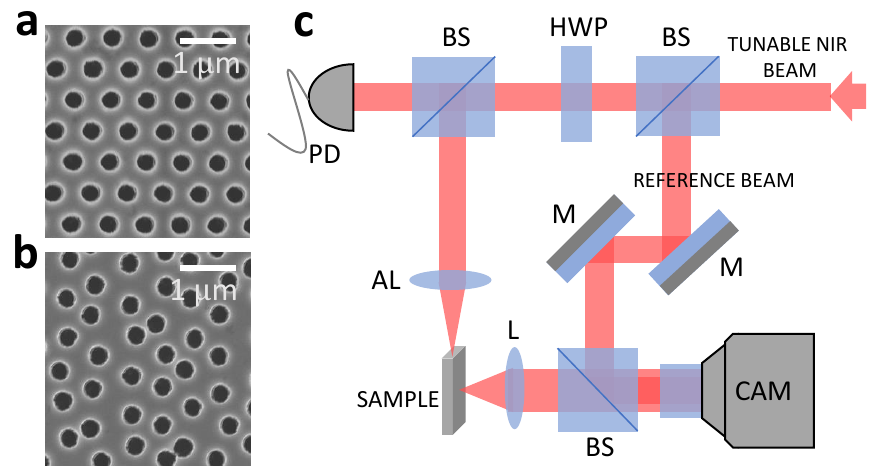}
	\caption{\label{fig:fig4} (a) and (b): SEM images of a crystalline and a disordered sample. The scalebar is 1 $\mu$m. (c): Experimental setup used in this study. HWP: Half wave plate; BS: Beam splitter; PD: Photo detector; M: Mirror; AL: Aspheric lens; L: Lens; CAM: Infrared camera.}
\end{figure}

Denoting the object beam and the reference beam at the CCD sensor plane by $O(x,y)$ and $R(x,y)$ respectively, the interference pattern $g(x,y)$ recorded by the sensor may in general be described by:
\begin{equation}
	g=|O|^2+|R|^2+(O^*R+OR^*).
\end{equation}
The problem of interest to us is to estimate the phase map $\phi_O(x,y)$ associated with the object beam $O(x,y)$. To this end, we first subtracted the individually recorded object and reference beam intensities $|O(x,y)|^2$ and $|R(x,y)|^2$ from the interference record $g(x,y)$ to retain information about the cross terms alone as follows:
\begin{equation}\label{eq:subtract}
h(x,y) = g(x,y) - |R(x,y)|^2 - |O(x,y)|^2.
\end{equation} 
Since the reference wave is a tilted plane wave as illustrated in Fig. 1, the phase map of the reference wave $R(x,y)$ has the typical form $\phi_R(x,y)=2\pi(f_{0x}x+f_{0y}y)$.  A two-dimensional Fourier transform of $h(x,y)$ is observed to have two well-separated energy lobes that are centered on the carrier fringe frequencies $(\pm f_{0x}, \pm f_{0y})$. We estimated the  spatial frequencies $f_{0x}$ and $f_{0y}$ by localizing the centroids of the cross-term lobes in the 2-dimensional Fourier transform of $h(x,y)$ with sub-pixel accuracy  \cite{singh2016, goyal2024}. The next processing step involves low-pass filtering of the product $[h(x,y) \times R(x,y)]$ by a circle shaped filter in 2-dimensional Fourier transform domain (with a Hamming window for avoiding any ringing artifacts) in order to separate out the cross-term $h_1 (x,y) = |R|^2 O$.  The phase of $h_1(x,y)$ is the same as $\phi_O (x,y)$ that we wish to recover. Due to the imaging configuration, the interference pattern is available only over a small region of the sensor array and the subtraction step in Eq. (\ref{eq:subtract}) is essential in order to clearly identify the carrier fringe frequency. For more details see Supplementary Information (Section I) \cite{supplemetary}. 

\begin{figure}
	\includegraphics[width=8.5cm]{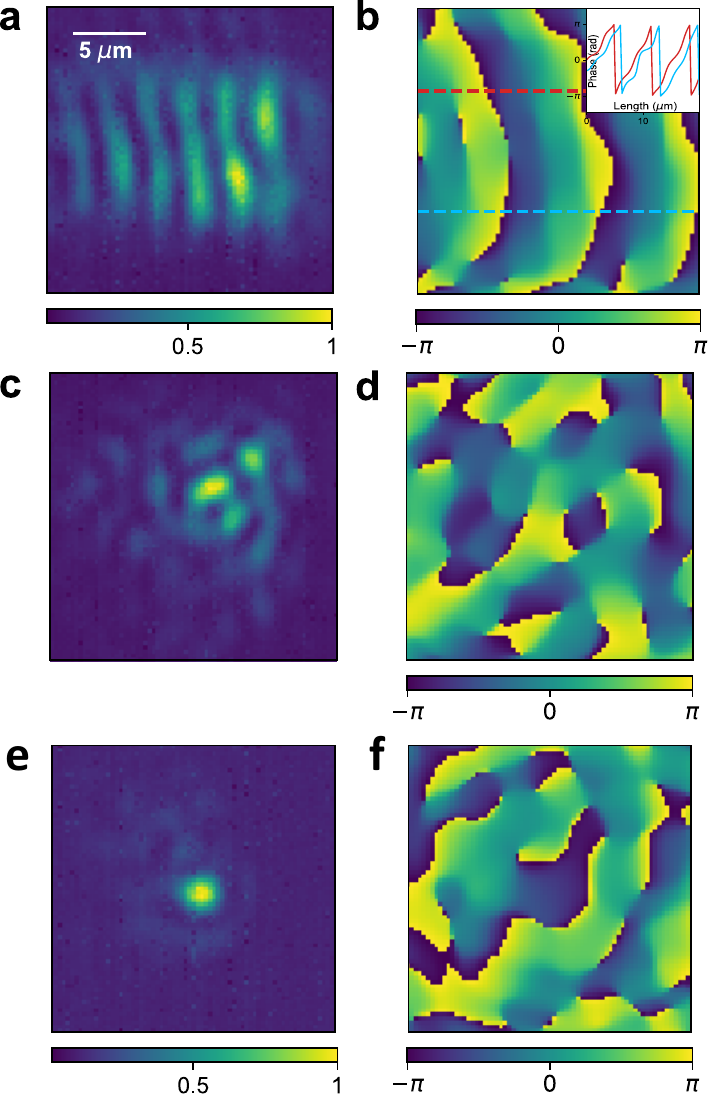}
	\caption{\label{fig:fig1} (a) Standing mode formed from counter-propagating (perturbed) Bloch waves in a nearly-periodic structure and (b) the extracted phase map therefrom. Inset shows cross-sections along the dashed red and blue lines. (c,d) and (e,f) show the same for weakly-localized and Anderson-localized modes.}
\end{figure}

The experimental measurements for a near-periodic sample are shown in Fig~2. Subplot (a) illustrates the standing mode formed from counter-propagating perturbed Bloch waves at $\lambda = 1575$~nm with the corresponding extracted phase-map $\phi_O(x,y)$, wrapped between $-\pi$ and $\pi$, shown in (b).  Linear cross-sections at two positions in the phase map (as marked with dashed lines) are plotted in the inset. Next, we show the data for disordered samples. Two representative modes, one weakly localized and the other strongly localized are depicted in images (c) and (e) respectively, with the corresponding extracted phase maps illustrated in (d) and (f) respectively. The randomness in the spatial phases is very apparent in the images. Given the intensity $I(x,y)$ and phase $\phi_O(x,y)$, the wavefunction $\psi(x,y)$ is reconstructed as $\sqrt{I}\exp [i\phi_O(x,y)]$. 
\begin{figure}
	\includegraphics[width=8cm]{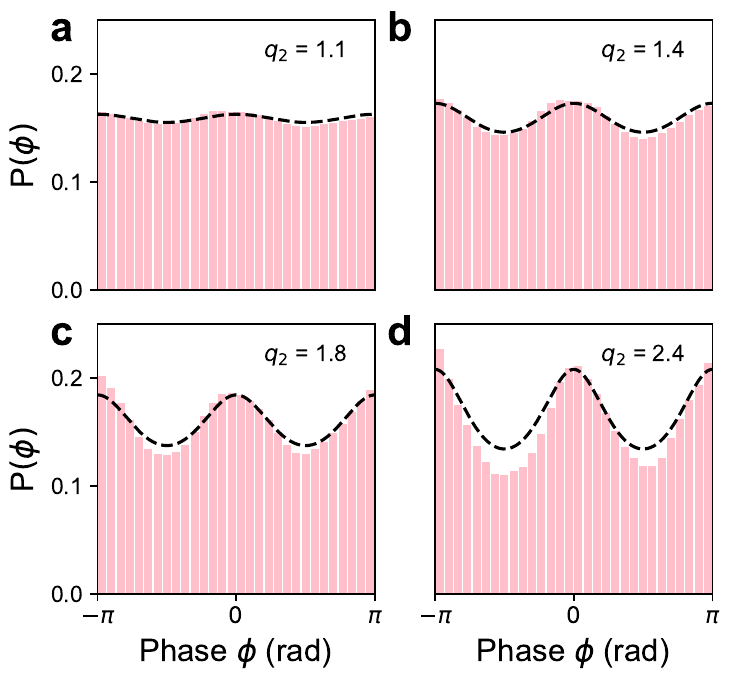}
	\caption{\label{fig:fig32} Measured phase distributions for varying degree of localization (histograms) compared with theoretical predictions (dashed lines) for four values of IPR $q_2$. Anderson localization leads to sharply peaked distributions about $0$ and $\pi$.}
\end{figure}

As a verification of the experimental procedure, the distribution of phases in localizing systems can be analysed, as they are a well-discussed feature in mesoscopic physics. Strongly localized modes are expected to exhibit a sharply peaked distribution of phase,  with the peaks at $0$ and $\pi$ rads. With decreasing localization, the peaks broaden out to yield a flat phase distribution \cite{vanneste2009}. We characterise  the spatial extents of modes in our system by their inverse participation ratio (IPR) $q_2 = N_p\Sigma_{i=1}^{N_p} I_i^2 / (\Sigma_{i=1}^{N_p} I_i)^2$, where $N_p$ is the the total number of pixels over which the mode is measured, and $I_i$ is the intensity at the $i^{\text{th}}$ pixel.  Depending on the spatial extent of a mode, $q_2$ ranges from 1 for an ideally extended mode to $N_p$ for a mode localized within a single pixel. 
Figure~3 displays the measured distribution of phases for four different $q_2$. Clearly, as the localization strengthens, the distribution peaks at $0$ and $\pi$ radians, in very good agreement with the theoretical predictions, as elaborated in SI \cite{supplemetary, lobkis00}.


\begin{figure}
	\includegraphics[width=8cm]{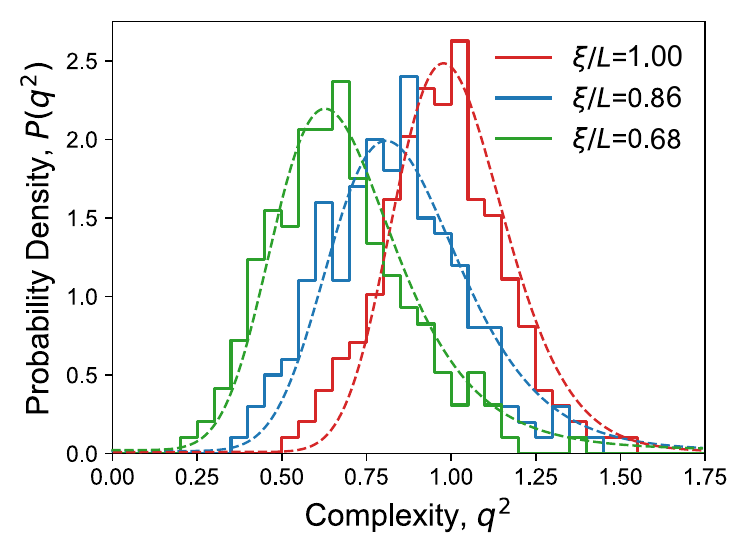}
	\caption{\label{fig:fig33} Distribution of complexities at different disorder strengths parameterized by the localization length $\xi$ (histograms). Dashed lines show theoretical fits (see Eq.\ (S14) of SI \cite{supplemetary}) with $\beta = \langle q^2 \rangle$ and $N$ as fit parameters ($\langle q^2 \rangle = 1$, 0.86, 0.68 and $N =$ 140, 68, 47 for the red, blue and green lines, respectively). }
\end{figure}

\begin{figure}
	\includegraphics[width=8cm]{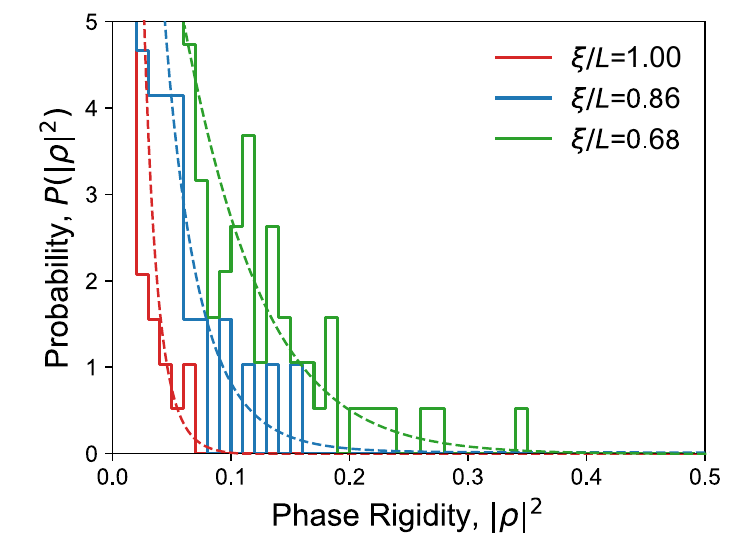}
	\caption{\label{fig:fig34} Probability distribution of phase rigidity $|\rho|^2$ for the same values of $\xi$ as in Fig.~5 (histograms). Dashed lines show theoretical predictions (see SI  Section II \cite{supplemetary}).}
\end{figure}

Subsequently, we addressed the behavior of the extracted complexities. To this end, we determined the mode localization length $\xi$ as an average over multiple modes  categorized as per their IPR $q_2$ into 10 bins. The mode profiles within each bin are averaged and an exponential decay is fit to the average profile. The complexity $q^2$ is also averaged over the same bins. Figure 4 illustrates the probability distribution of $q^2$ for modes with three different localization lengths $\xi$. For a nearly-periodic system (red histogram), the distribution is asymmetric and peaks around $q^2 = 1$. With decreasing $\xi$, the distribution shifts towards the lower values of $q^2$. Experimental data are fitted by a theoretical model that assumes $N \gg 1$ and treats both the numerator and the denominator of the definition of $q^2$ as Gaussian random variables \cite{supplemetary, hinkley69, kuethe00, marsaglia06, diaz13}. The first fit parameter of the model $\langle q^2 \rangle$ is discussed later, while the second fit parameter $N \sim 60$ estimates the effective number of independent terms in the definition of $q^2$, which should be of the order of the number of independent speckle spots in our measurements. In addition to inducing the complexity of the wavefunction, the openness of a disordered system also leads to fluctuations of the phase rigidity $|\rho|^2$. In open quantum systems, the probability distribution $P(|\rho|^2)$ determines the effective number of open channels through which the system is coupled to the environment \cite{brouwer2003}. In our disordered samples, $P(|\rho|^2)$ can be obtained directly from $P(q^2)$ by a simple transformation. The resulting expressions are quite lengthy \cite{supplemetary} but can be readily compared to the experimental data. Such a comparison is shown in Fig.~5. Note that the theory predicts an integrable divergence of $P(|\rho|^2) \propto 1/|\rho|$ for $|\rho|^2 \to 0$, in contrast to the predictions of finite $P(0)$ that exist for chaotic billiards \cite{langen1997,brouwer2003}.
 
Finally, we reveal a novel method to identify the localization length  by using the average complexity in the system.  Figure~6 discusses the variation of complexity with localization length. We first note that, as shown in the inset, the first fit parameter $\langle \beta \rangle$ of the above model nicely follows the expected dependence on $\xi/L$. The main plot depicts the variation of the measured $\langle q^2 \rangle$ with localization length. A theoretical model for this quantity can be obtained by assuming that in our disordered samples, $\langle q^2 \rangle$ is related to the typical mode width $\Gamma$ and the average mode spacing $\Delta$ in the same way as in an open chaotic system \cite{poli2009}:
$\langle q^2 \rangle = C \times \Gamma/\Delta$
where $C$ is a model-specific constant. The ratio $g =\Gamma /\Delta$ is known as the dimensionless conductance in the context of mesoscopic physics. In the regime of Anderson localization, it is $g \simeq \exp(-L/\xi)$. We thus have
$\langle q^2 \rangle \simeq \exp(c-L/\xi)$ with an unknown constant $c \sim 1$. Setting $c = 1$ yields a reasonable fit to the experimental data in Fig.~5. The fit is even better if we use a power series expansion of the above results around $\xi/L$ = 1: $\langle q^2 \rangle \simeq \xi/L$. The simple yet powerful relation $\langle q^2 \rangle = \exp(1-L/\xi) \simeq \xi/L$ for localized modes established above offers an immediate diagnostic into the localization parameters, such as $\xi$ or $g$. The extrapolation of this relation to tighter localization (not accessible experimentally in our samples) leads to $\langle q^2 \rangle = 0$ for $\xi = 0$, which carries a very plausible implication that a $\delta$-localized wavefunction should have no propagating component.

\begin{figure}
	\includegraphics[width=8cm]{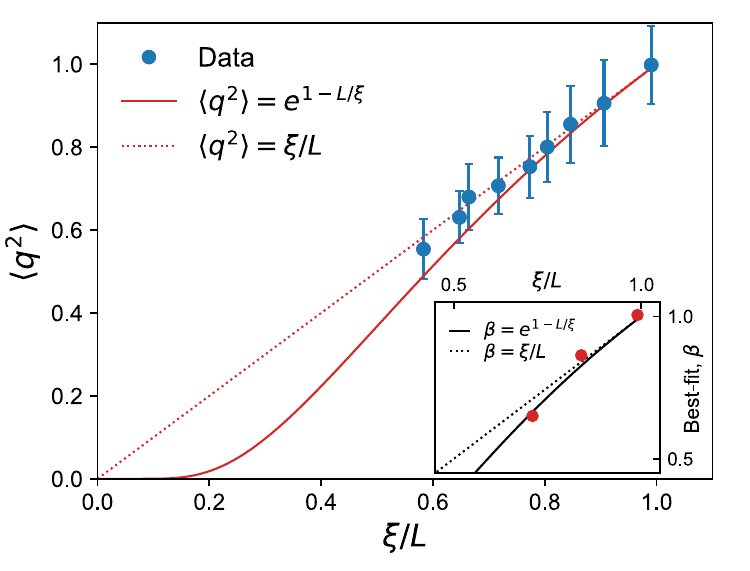}
	\caption{\label{fig:fig35}
 Average complexity as a function of the localization length $\xi$. A clear linear behavior is observed in the data. Solid and dotted lines show two theoretical predictions for $\langle q^2 \rangle$. Inset shows the best-fit values of $\beta = \langle q^2 \rangle$ obtained from the fits in Fig.\ \ref{fig:fig33} (symbols).}
\end{figure}

In conclusion, we present a novel criterion given by modal complexity to estimate the degree of light localization in disordered structures. We perform, to our knowledge, the first direct experimental measurement of the complexity of Anderson localized wavefunctions in open 2D disordered systems. Theoretical models developed for both the average complexity and its full statistical distribution allow us to estimate the localization length, thus demonstrating that complexity is an appropriate quantity for revealing and characterizing the Anderson localization phenomenon. Our work suggests a complexity-based metric of Anderson localization that may be an interesting alternative to other known criteria such as those proposed by Ioffe and Regel \cite{ioffe1960} or Thouless \cite{edwards1972,thouless1977}. Although we demonstrate its validity for localization of light in 2D, we believe that it can be readily applied to any type of waves in disordered media
because the theoretical arguments behind it are not specific for light or for two spatial dimensions. It would be particularly interesting to analyze complexity of localized modes in 3D experiments with ultrasound \cite{hu2008,cobus2018} and matter waves \cite{jendr2012,semeghini2015} as well as in recent  electromagnetic simulations \cite{yamilov2023}. Our work promises to reveal hitherto-unknown relations between complex fields and mesoscopic parameters of transport. It also emphasizes the need to address field statistics and not that of intensities, on which the current literature predominantly relies.


\subsection*{Acknowledgements}
We acknowledge expert sample fabrication by Monika Emmerling. We acknowledge helpful discussions with Kedar Damle, Rajdeep Sensarma and Vikram Tripathi. S. Mujumdar acknowledges the Swarnajayanti Fellowship from the Department of Science and Technology, Government of India.



\end{document}


\title{Supplementary Material for\\ ``Modal complexity as a metric for Anderson localization''}
\author{Sandip Mondal}
\affiliation{Nano-optics and Mesoscopic Optics Laboratory, Tata Institute of Fundamental Research, 1, Homi Bhabha Road, Mumbai, 400005, India}
\author{Kedar Khare}
\affiliation{Optics and Photonics Centre, Indian Institute of Technology Delhi, New Delhi 110016, India }
\author{Sergey E. Skipetrov}
\affiliation{Univ. Grenoble Alpes, CNRS, LPMMC, 38000 Grenoble, France}
\author{Martin Kamp}
\affiliation{Julius-Maximilians-Universit\"{a}t Würzburg, Physikalisches Institut, Lehrstuhl f\"{u}r Technische Physik, Am Hubland, 97074 W\"{u}rzburg, Deutschland}

\author{Sushil Mujumdar}
\email[]{mujumdar@tifr.res.in}
\homepage[]{http://www.tifr.res.in/~mujumdar}
\affiliation{Nano-optics and Mesoscopic Optics Laboratory, Tata Institute of Fundamental Research, 1, Homi Bhabha Road, Mumbai, 400005, India}

\maketitle
\section{Design of Disorder}
The primary periodic structure was designed for a hexagonal air-hole lattice (lattice constant $a=630nm$, hole radius $r=139nm$) in a GaAs membrane.  
Disordered samples were designed by displacing the center $(x, y)$ of each air-hole by uniformly distributed random numbers $(\partial x, \partial y)$, such that $\partial x^2+\partial y^2<d^2$, where $d=\frac{M}{2}\times\frac{\delta}{100}$. Here, $M=(a-2r)$ is the maximum possible displacement to avoid the coalescence of two holes. The disorder parameter $\delta$ was varied from a disorder of $0\%$ (periodic) to $100\%$, implying complete disorder, in steps of realizing structures from periodic, periodic-on-average random, to amorphous. Twenty-five configurations at each disorder strength were designed and fabricated. The membrane thickness was $340nm$, while the region area of the samples is $20\times20\mu m^2$.
\section{Phase Calculation}
In this Section, we describe the Fourier filtering methodology used for determining the phase map associated with object wave $O(x,y)$ as in Eq. (1) of the main manuscript. The interference pattern in Eq. (1) consists of four terms and the phase information of interest $\phi_o(x,y)=\arg[O(x,y)]$ is contained in the third and fourth cross terms ($O^* R+OR^*$). The phase extraction procedure for this off-axis interference configuration can be understood easily by referring to the 2D Fourier transform corresponding to the interference pattern $g(x,y)$. We denote the 2D Fourier transform of $g(x,y)$ as:
\begin{equation}
G(f_x,f_y )= \iint_{\mathbb{R}^2} dx\, dy\, g(x,y) \exp\left[-i 2\pi (f_x x+f_y y)\right].
\end{equation}

The 2D Fourier transform for discretely sampled images read out from the camera may be evaluated readily using the Fast Fourier Transform (FFT) routines available in commonly used programming frameworks such as MATLAB, GNU Octave or Python. The interference pattern $g(x,y)$ has four terms. It may be noted that the magnitude $|R|$ of the tilted plane reference wave (which may be recorded separately by blocking the object beam) is ideally constant across the array sensor. In practice, $|R|$ may at most show some low spatial frequency background-like features. The first two intensity terms $|O|^2$ and $|R|^2$ are located at the center ($f_x=f_y=0$) of the Fourier space whereas the cross-terms are centered on the carrier frequencies ($\pm f_{x0}$, $\pm f_{yo}$) associated with the reference beam $R$. This structure of the Fourier transform may be readily understood if we re-write the cross terms in the interference pattern in Eq. (1) as:
\begin{equation}
(O^* R+OR^* ) = 2 |R| |O| \cos\left[2\pi (f_{x0} x+f_{y0} y)-\phi_o(x,y)\right].
\end{equation}

A typical interference pattern $g(x,y)$ and its 2D Fourier transform magnitude $|G(f_x,f_y)|$ are shown in Fig. S1 (a) and (b) respectively. Since the recorded camera frame contains the interference pattern over a small area of the detector and the fringes have a moderate spatial frequency, we observe that the Fourier transforms of the first two intensity terms and the two cross terms in $g(x,y)$ are highly overlapped. It is therefore practically much easier to work with the function $h(x,y)$ in Eq. (2), which is obtained by subtracting the individually recorded intensity terms from the interference pattern. The function $h(x,y)$ and its Fourier transform magnitude $|H(f_x,f_y)|$ are shown in Fig. S1 (c) and (d) respectively. We observe that the energy lobes corresponding to the two cross-terms are now distinctly separated. The Fourier transform magnitudes are displayed with power of 0.25 for clarity so as to adjust for the dynamic range of the image display. The centroid of one of these energy lobes corresponds to the carrier fringe frequency $(f_{x0},f_{yo})$ associated with the tilted plane wave reference beam. The carrier frequencies may be estimated by first determining the centroids of the two energy lobes in the Fourier domain to integer pixel locations and further refining this estimate to sub-pixel accuracy by a local Discrete Fourier transform computation as per Ref. [27] of the manuscript. An erroneous determination of $(f_{x0},f_{yo})$ will lead to a low beat frequency background in the eventual amplitude and phase of the object wave $O(x,y)$. The next step is to multiply $h(x,y)$ by $R(x,y)=|R(x,y)| \exp[i 2\pi (f_{x0} x+f_{y0} y)]$ to obtain $h_1(x,y)$. The Fourier transform magnitude $|H_1(x,y)|$ is illustrated in Fig. S2 (a) which shows that the energy lobe corresponding to the cross term $R^* O$ is now shifted to the center of the Fourier space. A circle shaped mask may now be placed at the center of the Fourier space to filter out the term of interest to us as shown in Fig. S2 (b). An inverse Fourier transform operation after the filtering process leads to an estimate of the complex-valued function $|R|^2 O$ whose phase is the same as the phase map $\phi_o(x,y)$ corresponding to the unknown object function $O(x,y)$ that we wish to estimate. A 2D Hamming window has additionally been applied to the circular mask prior to performing the inverse Fourier transform operation in order to avoid ringing artefacts that may arise due to sharp edge of the circle shaped filter. The statistics of the numerical phase values of $\phi_o(x,y)$ at the individual pixels is of interest in this manuscript. For this purpose the phase values within the central dotted circle have been used. This is because the magnitude of the object wave $|O(x,y)|$ is negligible outside this circle making it meaningless to talk about their phase.
\begin{figure}[htbp]
\centering
\includegraphics[width=0.99\textwidth]{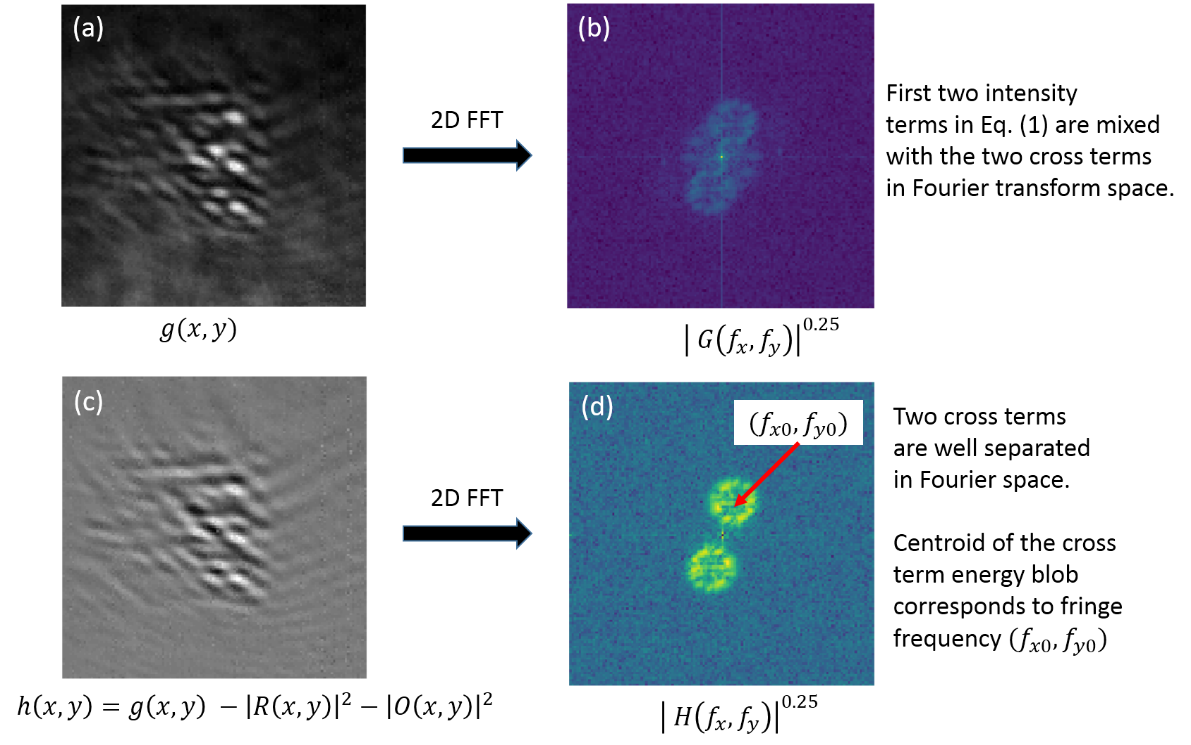}
\caption{Fig. S1: (a), (b) A typical interference pattern $g(x,y)$ and its 2D Fourier transform magnitude $|G(f_x,f_y)|$; (c), (d) Function $h(x,y)$ in Eq. (2) and its corresponding Fourier transform magnitude $|H(f_x,f_y)|$. The Fourier magnitudes are shown with a power 0.25 to adjust for the dynamic range of the image display.}
\end{figure}
\begin{figure}[htbp]
\centering
\includegraphics[width=0.99\textwidth]{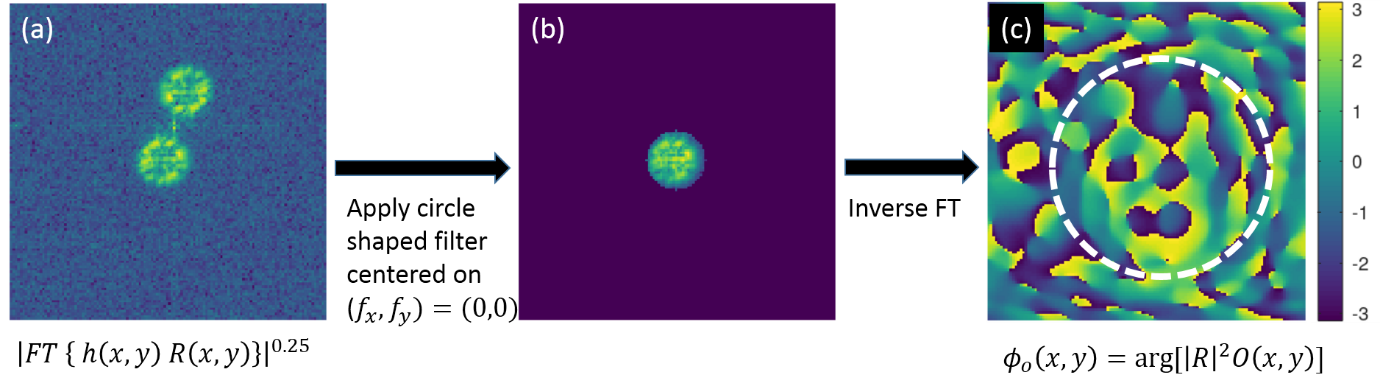}
\caption{Fig. S2: (a) Fourier transform magnitude corresponding to $h_1(x,y)=h(x,y)R(x,y)$ showing that the energy lobe corresponding to the cross term $R^* O$ is now shifted to the center of the Fourier space. (b) A circle shaped low-pass filter is applied to the Fourier transform in (a); (c) Phase of $|R|^2 O(x,y)$ estimated using the inverse Fourier transform applied to (b). The phase values within the dotted circle are used for statistics.}
\end{figure}

\section{Complexity and Phase Rigidity Distribution}
\subsection{Definition}
\label{def}

In the following, we use the following notations. The  complex wave function is
\begin{eqnarray}
\psi(\vec{r}) = e^{i \phi_0} \left[ \psi_r(\vec{r}) + i \psi_i(\vec{r}) \right]
\label{psi}
\end{eqnarray}
where  $\psi_r(\vec{r})$ and $\psi_i(\vec{r})$ are real-valued functions.

The complexity (or complexness) parameter is defined as
\begin{eqnarray}
q^2 = \frac{\frac{1}{N} \sum_{n=1}^N \psi_i(\vec{r_n})^2}{\frac{1}{N} \sum_{n=1}^N \psi_r(\vec{r_n})^2}
\label{complex}
\end{eqnarray}
where the sums are over $N$ pixels $n$ at which the field $\psi(\vec{r}_n)$ is measured.

The phase rigidity is 
\begin{eqnarray}
\left| \rho \right|^2 = \left( \frac{1-q^2}{1+q^2} \right)^2.
\label{rho2}
\end{eqnarray}

The inverse participation ratio (IPR) is
\begin{eqnarray}
q_2 = \frac{\frac{1}{N} \sum_{n=1}^N I(\vec{r_n})^2}{\left[\frac{1}{N} \sum_{n=1}^N I(\vec{r_n} ) \right]^2}.
\label{ipr}
\end{eqnarray}
For a mode that is essentially different from zero only on $M$ pixels, we have $I(\vec{r}) \sim 1/M$ and  
\begin{eqnarray}
q_2 \simeq \frac{N}{M} \simeq \left( \frac{L}{\xi} \right)^2 = \frac{1}{\langle q^2 \rangle^2}
\label{ipr2}
\end{eqnarray}
where we used the fact $N \propto L^2$ and $M \propto \xi^2$ for an isotropic system in 2D, as well as the relation $\langle q^2 \rangle = \xi/L$ derived in the main text.

\subsection{Distribution of the phase}
\label{phase}

The phase of a the complex field (\ref{psi}) can be defined as 
\begin{eqnarray}
\phi(\vec{r}) = \text{arctan} \left[ \frac{\psi_i(\vec{r})}{\psi_r(\vec{r})} \right].
\label{phi}
\end{eqnarray}
The statistical distribution of $\phi$ has been derived by Lobkis ans Weaver \cite{lobkis00} assuming that  $\psi_r$ and $\psi_i$ are independent Gaussian random variables with a zero mean:
 \begin{eqnarray}
P(\phi) = \frac{\tilde{q}}{2\pi} \times \frac{1}{\tilde{q}^2 \cos^2 \phi + \sin^2 \phi}
\label{pphi}
\end{eqnarray}
where
\begin{eqnarray}
\tilde{q}^2 = \frac{\langle \psi_i(\vec{r})^2 \rangle}{\langle \psi_r(\vec{r})^2 \rangle}.
\label{qtilde}
\end{eqnarray}
Note that, strictly speaking, $\tilde{q}^2 \ne  \langle q^2 \rangle$ because of the different ways of averaging. If we neglect this subtlety, we have $\tilde{q} =  \langle q^2 \rangle^{1/2} \simeq q_2^{-1/4}$. This yields the theoretical predictions for $P(\phi)$ shown by dashed lines in Fig.\ 4 of the main text.

\subsection{Probability distribution of complexity}
\label{complexity}

To evaluate the statistical distribution of $q^2$, we note that both the numerator
$X = (1/N) \sum_{n=1}^N \psi_i(\vec{r_n})^2$
and the denominator
$Y = (1/N) \sum_{n=1}^N \psi_r(\vec{r_n})^2$
of Eq.\ (\ref{complex}) are sums of $N \gg 1$ random variables. If we assume that the latter are independent, then the central limit theorem states that both $X$ and $Y$ should have Gaussian statistics with means and variances
\begin{eqnarray}
\mu_X &=& \langle X \rangle = \frac{1}{N} \sum_{n=1}^N   \left\langle \psi_i(\vec{r_n})^2 \right\rangle = \left\langle \psi_i(\vec{r_n})^2 \right\rangle = \langle I_i \rangle
\;\;\;\;\;\;\;\;
\label{mux}
\\
\mu_Y &=& \langle Y \rangle = \frac{1}{N} \sum_{n=1}^N   \left\langle \psi_r(\vec{r_n})^2 \right\rangle = \left\langle \psi_r(\vec{r_n})^2 \right\rangle = \langle I_r \rangle
\label{muy}
\\
\sigma_X^2 &=& \frac{\sigma_i^2}{N},\;\;\;\;
\sigma_Y^2 = \frac{\sigma_r^2}{N}
\label{sxsy}
\end{eqnarray}
where we denote $I_{r,i}(\vec{r}) = \psi_{r,i}(\vec{r_n})^2$  and $\langle I_{r,i} \rangle = \langle \psi_{r,i}(\vec{r_n})^2 \rangle$,
$\sigma_{r,i}^2 = \langle [I_{r,i}(\vec{r}) -   \langle I_{r,i} \rangle]^2 \rangle$.
Note that because we assume that $N$ terms in sums of Eq.\ (\ref{complex}) are statistically independent, $N$ will correspond to the effective number of independent speckle spots in the field of view of the measurement device and not to the number of points (pixels) at which the electromagnetic field is measured.

Probability density of a ratio $Z = X/Y$ of two Gaussian random variables can be derived using the general approach first used by Hinkley \cite{hinkley69}. Explicit formulas have been given in several papers \cite{kuethe00,marsaglia06,diaz13}; we use an expression from Ref.\  \cite{diaz13}:
\begin{eqnarray}
p_Z(z) &=& \frac{\rho}{\pi (1 + \rho^2 z^2)}
\exp \left[ -\frac{\rho^2 \beta^2 + 1}{2 \delta_y^2} \right]
\nonumber\\
&\times&
\left\{
1 + \sqrt{\frac{\pi}{2}} \;Q\; \text{erf} \left( \frac{Q}{\sqrt{2}} \right)
\exp\left( \frac{Q^2}{2} \right)
\right\}
\label{pz}
\end{eqnarray}
where
\begin{eqnarray}
Q &=& \frac{1 + \beta \rho^2 z}{\delta_y \sqrt{1 + \rho^2 z^2}}
\label{q}
\end{eqnarray}
and $\beta = \mu_x/ \mu_y$, $\rho = \sigma_y/\sigma_x$, $\delta_y = \sigma_y/\mu_y$.
For $N \gg 1$, we find $\langle q^2 \rangle = \beta$ to a good accuracy. 

We assume that $\sigma_{i}/\sigma_r = \mu_i/\mu_r$ typical for light scattering in diffuse regime (corrections due to Anderson localization amplify both $\sigma_{i}$ and $\sigma_{r}$ but preserve their ratio equal to $\mu_i/\mu_r$) and treat $\beta = \langle q^2 \rangle$ and $N$ as fit parameters to adjust Eq.\ (\ref{pz}) to experimental data. The results are shown in Fig.~6 of the main text by dashed lines. The parameter $\beta$ controls the position of the maximum of the distribution $P(q^2)$ and its mean whereas $N$ determines the width of the distribution.

\subsection{Probability distribution of phase rigidity}

To shorten the notation, let us define $x = q^2$ and $y = |\rho|^2$. Then
\begin{eqnarray}
y(x) &=& \left( \frac{1-x}{1+x} \right)^2
\label{yx}
\\
x_{1,2}(y) &=& \frac{1 \pm \sqrt{y}}{1 \mp \sqrt{y}}
\label{xy}
\\
p_y(y) &=& \sum\limits_{n=1}^2 p_x[x_n(y)]  \left| \frac{d x_n(y)}{dy} \right|
\nonumber \\
&=&
\sum\limits_{n=1}^2 p_Z\left( \frac{1 \pm \sqrt{y}}{1 \mp \sqrt{y}} \right)  
\frac{1}{(1 \mp \sqrt{y})^2 \sqrt{y}}
\label{pxy}
\end{eqnarray}
where we substituted $p_Z(x)$ for $p_x(x)$ and used
\begin{eqnarray}
\left| \frac{d x_n(y)}{dy} \right|
&=& \frac{1}{(1 \mp \sqrt{y})^2 \sqrt{y}}.
\label{der}
\end{eqnarray}
Equation (\ref{pxy}) is shown by dashed lines in Fig.~7 of the main text for the same values of $\langle q^2 \rangle$ and $N$ as those in Fig.~6.

%